\documentclass[prl,aps,twocolumn,superscriptaddress,floatfix,showpacs]{revtex4}
\usepackage{bm}
\usepackage{amssymb}
\usepackage{graphicx}

\begin{document}

\title{Helicity, anisotropies and their competition in a multiferroic magnet: \\
insight from the phase diagram}

\author{M. V. Gvozdikova}
\affiliation{%
Institut Laue Langevin, Bo\^ite Postale 156, F-38042 Grenoble Cedex 9, France}
\author{T. Ziman}
\affiliation{%
Institut Laue Langevin, Bo\^ite Postale 156, F-38042 Grenoble Cedex 9, France}
\affiliation{%
LPMMC, UMR-5493, Universit\'e Grenoble Alpes and CNRS, 38042 Grenoble, France}
\author{M. E. Zhitomirsky}
\affiliation{CEA, INAC-PHELIQS, F-38000, Grenoble, France}

\begin{abstract}
Motivated by the complex phase diagram of MnWO$_4$, we investigate the competition
between anisotropy, magnetic field, and helicity for the anisotropic next-nearest-neighbor
Heisenberg model. Apart from two competing exchanges, which favor a spiral
magnetic structure, the model features the bi-axial single-ion anisotropy. The model
is treated in the real-space mean-field approximation and the phase diagram containing
various incommensurate and commensurate states is obtained for different field orientations.
We discuss the similarities and differences of the theoretical phase diagram and the experimental diagram of
MnWO$_4$.
\end{abstract}

\date{June 8, 2016}%


\maketitle


{\it Introduction.}---%
Phase diagrams of magnetic materials contain important information about their
atomic-scale interactions. Competition between exchange interactions and anisotropy
may produce nearly degenerate states that exhibit remarkable sensitivity to an applied
magnetic field leading to rich and complex phase diagrams. Such a situation is often
realized in spiral multiferroics, where helicity results from frustrated exchanges and
a sizable spin-orbit interaction is a source of coupling between local magnetization
and electric polarization \cite{Cheong07,Tokura14}. An incomplete list of multiferroic
materials with numerous incommensurate and commensurate magnetic states includes
TbMnO$_3$ \cite{Kajimoto04,Kenzelman05},
Ni$_3$V$_2$O$_8$
\cite{Lawes04,Lawes05,Ehlers13}, CeFeO$_2$ \cite{Kimura06}, CuO \cite{Yang89,Kimura08},
RbFe(MoO$_4$)$_2$ \cite{Kenzelman07},  and MnWO$_4$
\cite{Lauten93,Arkenbout06,Heyer06}.

Recently, significant progress was made in the reconstruction of the full phase diagram
of MnWO$_4$ with the help of  neutron diffraction \cite{Nojiri11} and  electric
polarization \cite{Mitamura12} measurements in pulsed magnetic fields. Despite
substantial experimental \cite{Ehrenberg99,Ye11,Urcelay14} and theoretical
\cite{Tian09,Solovyev13,Toledano10,Sakhnenko10,Matityahu12,Quirion13} efforts a full
explanation of the complex phase diagram of MnWO$_4$ is still lacking. Here we adopt
a strategy different from the phenomenological theories of MnWO$_4$
\cite{Toledano10,Sakhnenko10,Matityahu12,Quirion13} by formulating and studying a minimal
spin model relevant to this magnetic material. The Landau energy functional for competing
multi-component order parameters typically has a large number of unknown
phenomenological parameters producing a significant degree of arbitrariness. Besides,
the Landau theory is not applicable at low temperatures and strong magnetic fields, where
interesting phase transformations take place. In contrast, the minimal spin model
contains the least possible number of coupling constants and can be simulated without
any {\it ad-hoc} assumption on equilibrium magnetic states.

In this work we investigate the anisotropic next-nearest-neighbor Heisenberg (ANNNH) spin
model. In addition to competing exchange interactions the model features a {\it bi-axial}
single-ion anisotropy, which is consistent with the monoclinic symmetry of MnWO$_4$. Basically,
this model is a generalization of the celebrated ANNNI model \cite{Fisher80,Bak80} to
three-component quantum spins. We obtain the $H$--$T$ phase diagram of the ANNNH model
using unrestricted real-space mean-field simulations. This approach has certain advantages
in comparison to the classical Monte Carlo simulations used before for spiral multiferroics
\cite{Furukawa09,Fishman12} as it includes local quantum fluctuations and allows us to predict
field and temperature variations of the ordering wave vectors (see details below). Our study
suggests that the field-induced transition into the commensurate state in MnWO$_4$ can be
produced by the bi-axial anisotropy, whose role in this was so far overlooked in the literature.
The topology  of the phase diagram of  MnWO$_4$ for magnetic fields along the easy axis is
perfectly reproduced within the ANNNH model.

The spin Hamiltonian of the model
\begin{eqnarray}
\hat{\cal H} & = & \hat{\cal H}_{ex} + \hat{\cal H}_{SI} \,,\ \
\hat{\cal H}_{ex}=\sum_{\langle ij\rangle} J_{ij} {\bf S}_i\cdot{\bf S}_{j}\,,
\nonumber \\
&& \hat{\cal H}_{SI} = \sum_i
\Bigl\{E[(S_i^x)^2\! - (S_i^y)^2]-D(S_i^z)^2\Bigr\}
\label{H}
\end{eqnarray}
describes an array of antiferromagnetic spin chains along the $c$-axis with competing first
$J_1$ and second $J_2$ neighbor exchange interactions. Coupling between chains in the $ab$
plane is assumed to be ferromagnetic $J_0<0$. The bi-axial single-ion anisotropy has
the easy axis along $z$ and the hard axis along $x$: $D>E>0$. Note, that
in low-symmetry crystals, orientation of the principal spin axes may differ from
the crystallographic directions.
For $J_2>J_1/4$ and weak anisotropy, the model has a spiral magnetic ground state
with the wave vector $\cos Q = -J_1/(4J_2)$ along the chain direction.
This toy model is often invoked for a description of real spiral antiferromagnets \cite{Elliot61,Johnston12,Nagamiya67,Lawes04,Harris07,Harris91}.

Generally, a weak easy-axis anisotropy splits a single transition temperature
of an exchange spiral antiferromagnet
into two separate transitions for longitudinal (higher $T_c$) and transverse (lower $T_c$) spin components \cite{Nagamiya67}.
In addition, MnWO$_4$ features the third low-temperature transition into
a commensurate collinear state with moments parallel to the easy axis.
The extra transition appears because an exchange energy loss in the commensurate state is surpassed by a gain in the anisotropy term.
In particular, this requires close values for  the wave vectors in the two magnetic structures.
To model such a situation in the framework of the ANNNH model, we fix $J_2/J_1=2$, which yields the spiral wave vector
$Q_{IC}/(2\pi)= 0.27$ close to the commensurate value $Q_C/(2\pi)=0.25$.

{\it Theory.}---%
To find possible ordered states of  the ANNNH model in an external magnetic field
$\bf H$ we use the real-space mean-field approach; see, for example, \cite{Melchy09,Chaudhury08}.
The mean-field theory begins with defining local averages ${\bf m}_i =\langle {\bf S}_i\rangle$
and neglecting intersite correlations  $\langle({\bf S}_i-{\bf m}_i)({\bf S}_j-{\bf m}_j)\rangle=0$
in the exchange term. In the mean-field approximation, the spin Hamiltonian transforms into
\begin{equation}
\hat{\cal H}_{MF} = \hat{\cal H}_{SI} - \sum_i{\bf h}_i\cdot{\bf S}_i
- \sum_{\langle ij\rangle} J_{ij} {\bf m}_i\cdot{\bf m}_{j}   \,,
\label{HMF}
\end{equation}
where the local fields are  ${\bf h}_i = {\bf H} - \sum_j  J_{ij} {\bf m}_j$.
Because of the single-ion anisotropy, the dependence of ${\bf m}_i$
on ${\bf h}_i$ is not described by the Brillouin function.
Instead, we have diagonalized the local Hamiltonian matrix for
a given ${\bf h}_i$ and $S=5/2$ (assuming Mn$^{2+}$ ions)
and computed ${\bf m}_i$ numerically. The mean-field Hamiltonian (\ref{HMF}) has been simulated
on finite clusters with periodic boundary conditions.
To match the incommensurate wave vector $Q_{IC}$ the linear dimension  along chains has to be
chosen at least $L=100$ sites. On the other hand,  the commensurate  $Q_\perp=0$ ($J_0<0$)
allows us to consider in the mean-field approximation only a single chain, replacing the effect of neighboring
chains by an effective field.

\begin{figure}[t]
\centerline{
\includegraphics[width=0.75\columnwidth]{OPvsT.eps}
}
\vskip 3mm
\centerline{
\includegraphics[width=0.77\columnwidth]{OPvsH.eps}
}
\vskip 4mm
\centerline{
\includegraphics[width=0.99\columnwidth]{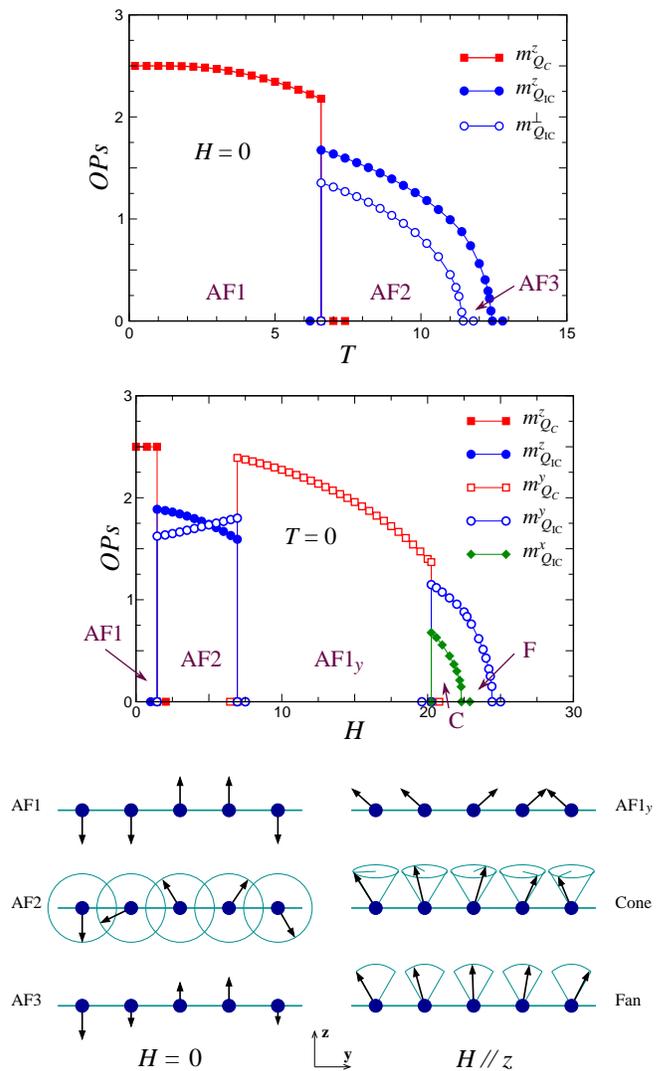}
}
\caption{Upper panel: temperature dependence of the order parameters in zero field for the
uniaxial anisotropy: $D=0.2$ and $E=0$. Middle panel:
field  dependence of the order parameters at $T=0$ in the bi-axial case:  $D=0.3$ and $E=0.1$.
Lower panel: sketch of the six states of the ANNNH model in magnetic field ${\bf H}\parallel z$
with corresponding labels. For illustration purpose, the easy $z$ axis  is chosen to be orthogonal to the
chain $c$ direction.
}
\label{fig:OPs}
\end{figure}

For fixed $H$ and $T$, we start with a random set of $\{{\bf m}_i\}$ and iterate repeatedly
the self-consistency condition for all sites until convergence. The procedure is performed
for up to $10^3$ initial random configurations and a solution with the lowest free-energy is
selected. For the obtained spin structure we calculate the Fourier harmonics $m^\alpha_{q}$
for all possible wave vectors $q=2\pi n/L$ with integer $n$ and pick up
the maximum amplitude for each $\alpha$.

{\it Results.}---%
Let us begin with the behavior in zero magnetic field.
We have performed the real-space mean-field simulations of the model (\ref{H}) with
$J_1=1$, $J_2=2$ and various values for $D$, $E$, and  $J_0$.
The role of interchain coupling $|J_0|\leq J_1$ consists, for the most part, of a trivial shift in all characteristic
temperatures by $\Delta T = zS(S+1)|J_0|/3$, where $z$ is the number of
nearest-neighbor chains.  For brevity we show only the results obtained with $J_0=0$.
The typical behavior for a moderate anisotropy  $D=0.2$ is shown in
the top panel of Fig.~\ref{fig:OPs}. We use the standard convention adopted for MnWO$_4$ and
label the ordered antiferromagnetic phases from low to high temperatures as AF1, AF2, and AF3.
The corresponding spin structures are sketched in the bottom panel of Fig.~\ref{fig:OPs}.
The collinear AF1 state described by the commensurate wave vector $Q_C/2\pi= 0.25$
is stable below $T_{c1}\approx 6.6$.
The elliptical spiral AF2 state  exists at $T_{c1}<T<T_{c2}\approx 11.4$ and the collinear sinusoidal
AF3 state appears at $T_{c2}<T<T_{c3}\approx 12.4$ with the incommensurate  propagation vector
$Q_{IC}/2\pi\approx 0.27$ in both cases.

Three successive transitions are present for $0.17\alt D \alt 0.35$.
For smaller anisotropy, $D\leq 0.15$, the model exhibits only two
transitions with the elliptical spiral state  stable for all
$T< T_{c2}$. For larger anisotropy,
$D\geq 0.4$, the spiral phase disappears, opening up a direct
transition between collinear commensurate and incommensurate states. Such a behavior is observed
in iron-doped Mn$_{1-x}$Fe$_x$WO$_4$ for $x\geq 5$~\%, where Fe$^{2+}$ ions are believed to enhance the local anisotropy
\cite{Chaudhury08,Ye08}.

Even though the zero-field  behavior of MnWO$_4$ can be satisfactorily accounted for by a uniaxial anisotropy,
theoretical description of field-induced states requires us to include an in-plane term $E$.
The middle panel of Fig.~\ref{fig:OPs} shows the field evolution of order parameters at $T=0$ for $D=0.3$ and $E=0.1$
with the field applied along the easy axis. The magnetization process features five distinct antiferromagnetic phases before
transition into the saturated state at $H_s\approx 24$. Apart from the common conical (C) and fan (F) magnetic structures \cite{Nagamiya67},
there is a wide region of the commensurate antiferromagnetic state with a nonzero $m^y_{Q_C}$, which we accordingly denote as the AF1$_y$ state.

Essentially the sequence of ordered states, F$\to$C$\to$ AF1$_y$, upon decreasing magnetic field at $T=0$
repeats the sequence AF3$\to$AF2$\to$AF1 upon cooling in zero field with active spin components
rotating in the $xy$ and the $yz$ planes, respectively. In particular, presence of the commensurate AF1$_y$ state
requires a substantial difference between the intermediate $y$ axis and the hard $x$ axis
to compensate the exchange energy loss with respect to the incommensurate conical structure.
The field region occupied by the AF1$_y$ state shrinks for decreasing $E$ 
and completely goes away for $E\alt 0.08$. In turn, the conical state disappears for $E\agt 0.13$ opening a direct F$\to$AF1$_y$ transition.
Note, that the distorted conical and the fan states both have a small longitudinal harmonic
$m^z_{2Q_{\rm IC}}$, which is a subdominant order parameter and, therefore, not included in Fig.~\ref{fig:OPs}.

\begin{figure}[t]
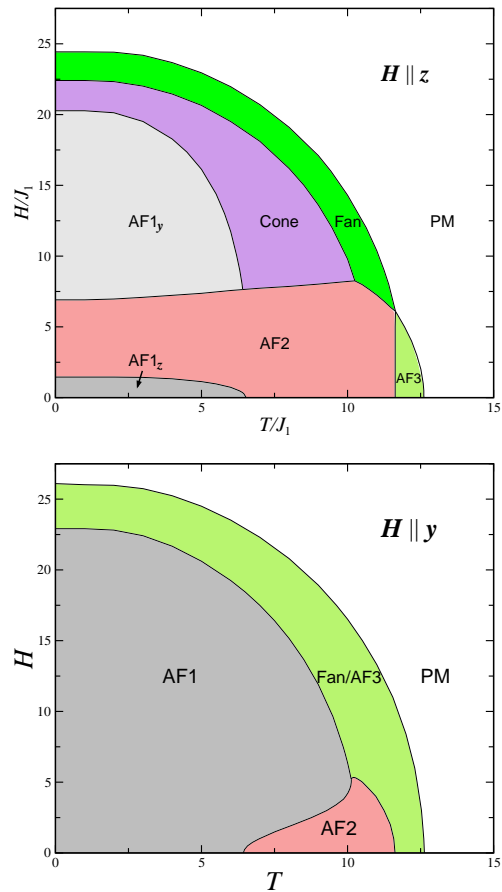

\centerline{
\includegraphics[width=0.75\columnwidth]{PhDz.eps}
}
\vspace{3mm}
\centerline{
\includegraphics[width=0.75\columnwidth]{PhDy.eps}
}
\caption{The $H$--$T$ phase diagrams for magnetic field along the easy $z$ axis (upper panel) and
along the intermediate $y$ axis (lower panel). The single-ion anisotropy constants are $D=0.3$ and $E=0.1$.
}
\label{fig:PhD}
\end{figure}

The $H$--$T$ phase diagram of the ANNNH model for the field parallel to the $z$ axis
is shown in the upper panel of Fig.~\ref{fig:PhD}. The obtained diagram is strikingly similar
to the experimental phase diagram of MnWO$_4$ for fields along the easy direction \cite{Nojiri11,Mitamura12}.
Since the experiments were performed in  pulsed  magnetic fields $H\simeq 30$--50~T, only the low-field states
of MnWO$_4$ were fully characterized so far. Our theory strongly suggests  that the experimental states IV and V have
the conical and the fan structure, respectively. Accordingly, the magnetoelectric effect was found only in
the IV (C)  state \cite{Mitamura12}. The field-induced commensurate state HF \cite{Nojiri11} is identified
with the AF1$_y$ phase with moments alternating along the intermediate $y$ axis,
which coincides with the two-fold crystallographic  $b$ axis. This finding fully agrees with the recent
optical absorption measurements \cite{Toyoda15} and with the phenomenological theory \cite{Quirion13}.

The phase diagram of the ANNNH model for magnetic field applied along the $y$ axis is shown
in the lower  panel of Fig.~\ref{fig:PhD}. The commensurate AF1 state occupies a significant
part of the ordered region. In high magnetic fields the AF1 state is succeeded by the fan state.
Since the spin polarization in the fan structure for ${\bf H}\parallel y$ is the same as in the AF3 state in zero field,
the two phases are described by the same order parameter $m^z_{Q_{IC}}$ and continuously transform into each other.
Overall, the theoretical diagram closely resembles the experimental diagram for ${\bf H}\parallel b$ \cite{Mitamura12}.
The only difference between the two is a narrow strip of the magnetoelectric $X$ phase between the AF1 and fan states present
in MnWO$_4$ \cite{Urcelay14,Mitamura12}. This phase has a distorted cycloidal (conical) order in the $ac$ ($xz$) plane
and appears in our simulations for smaller values of $D$.
The selected anisotropy parameters are, however, fixed to mimic the experimental ratio $T_{c1}/T_{c3}\approx 0.5$ in  zero field.
In order to fully reproduce the phase diagram of MnWO$_4$  for ${\bf H}\parallel b$ one no doubt has
to consider a more realistic pattern of exchange interactions that would allow appropriate modification
of anisotropy constants.

Finally, we have studied the temperature and field variations of the ordering wave vector.
The published experimental data for MnWO$_4$ indicate close but distinct propagation vectors
for the incommensurate states in zero field and above $H=10$~T \cite{Nojiri11,Urcelay14}.
Continuous variations of the ordering wave vector were also observed for TbMnO$_3$ \cite{Kenzelman05} and RbFe(MoO$_4$)$_2$ \cite{Mitamura14}.
From a theoretical perspective, the problem is quite challenging because discreteness of the wave vectors for
a single cluster inevitably produces spurious phase transitions related to the propagation vector jumps.
Instead, we have simulated a range of clusters with different linear sizes $L$ selecting among them the magnetic structure
with the lowest free energy. About 120 clusters  with $10\leq L\leq 170$
were typically investigated for each $T$ and $H$.
A similar approach, albeit on a lesser scale, was used previously for the ANNNI model \cite{Bak80}.

\begin{figure}[t]
\centerline{
\includegraphics[width=0.85\columnwidth]{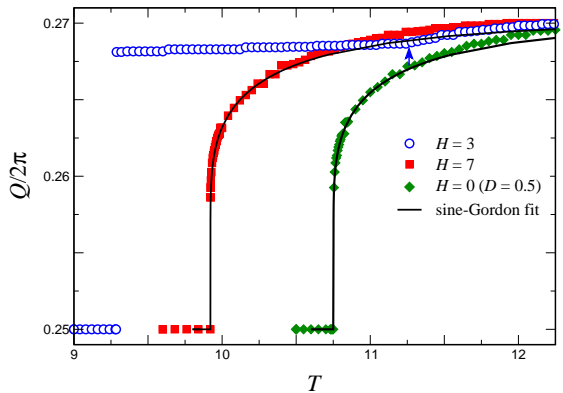}
}
\caption{Wave vector of the equilibrium magnetic structure versus temperature for $H=3$ and 7
along the $y$ axis ($D=0.3$ and $E=0.1$) and in zero field for $D=0.5$ and $E=0$.
The solid line gives sine-Gordon fits for the direct commensurate-incommensurate transition.
A small vertical arrow indicates the AF2-AF3 transition for $H=3$.
}
\label{fig:QT}
\end{figure}

Figure \ref{fig:QT} shows the temperature dependence of the ordering wave vector for magnetic fields parallel to
the  $y$ axis. For the $H=3$ scan, the propagation vector exhibits a  jump at the
AF1--AF2 boundary accompanied by smooth weak variations inside the AF2 and AF3 phases.
In zero field (not shown), $Q$ changes even less, by only  0.5\%\  between  $T_{c1}$ and $T_{c3}$.
The behavior becomes notably different for scans that cross the AF1--AF3 boundary.
Rapid variations of $Q$ are clearly seen for $H=7$  and also
in zero field, once a strong easy-axis anisotropy
($D=0.5$) suppresses the spiral phase.

The behavior of $Q$ near the AF1--AF3 boundary
can be interpreted as follows. The ordered spin components
in both states are parallel to the $z$ axis.  Consequently,
the free energy for the corresponding transition is expressed as a function of
a complex scalar order parameter $m^z_{Q_C}({\bf r})$, which is uniform in the commensurate AF1 state
and acquires a position dependent phase $m^z_{Q_C}({\bf r}) \sim e^{i\phi({\bf r})}$ in the incommensurate AF3 state.
In the constant amplitude approximation assuming slow $z$ variations, the free energy acquires the form
\begin{equation}
F = \int dz \Bigl[ \frac{K}{2} \Bigl(\frac{d\phi}{dz} - \delta\Bigr)^2 + \frac{V}{4} \cos 4\phi\Bigr]\,,
\label{F}
\end{equation}
where $\delta = Q_{IC}-Q_C \approx J_1/4J_2$ and $V\propto T$ \cite{Bak80}.
In the equilibrium  state, $\phi({\bf r})$ satisfies the sine-Gordon equation
$\phi''_{zz}+(V/K)\sin4\phi =0$, which provides the basis for the analytic theory of the
$C$--$IC$ transition \cite{Dzyal65,Chaikin}. Changes in the propagation vector
$Q$ are attributed to the varying distance between solitons in a periodic soliton lattice.
The corresponding predictions are shown in Fig.~\ref{fig:QT} by solid lines. The excellent agreement between numerical
results and the analytic theory worsens towards the N\'eel temperature, signifying  departure from the simple $V\propto T$ law.
Interestingly, there is no sign of the devil's staircase in the temperature dependence of $Q$, which is known to exist for the closely related ANNNI model
\cite{Fisher80,Bak80}. The difference in the behavior between the two models can be related to quantum effects
present in the ANNNH model and deserves further investigation.

The close resemblance of the experimental and theoretical phase diagrams
suggests that the behavior of MnWO$_4$ in an external field is governed by competition between
helicity and the bi-axial anisotropy being essentially magnetic in nature. There
is no need to invoke other terms, such as a biquadratic exchange, which was suggested to play a role for $\rm Ni_3V_2O_8$ \cite{Ehlers13}.
The ferroelectricity appears as a secondary effect fully consistent with the
spin current mechanism \cite{Katsura07} with only the AF2 and the conical state  showing electric polarization.
The existence of the fan phase between conical and paramagnetic states confirms the observation
of a non-ferroelectric magnetic phase at high fields \cite{Mitamura12}. It would be also interesting
to confirm experimentally the multicritical point between fan, AF2 and AF3 phases predicted for ${\bf H}\parallel z$
in the present calculations and in the Landau theory \cite{Quirion13}.

Our study of the ANNNH model opens the door for a detailed theory of MnWO$_4$  using the multiple exchange constants deduced
from high resolution inelastic neutron scattering, see, e.g., \cite{Ye11}. Note that the presence of long-distance exchanges in
MnWO$_4$ improves the accuracy of the mean-field calculation for thermodynamic properties.
The real-space mean-field simulations can be also applied to other multiferroic materials with complex phase diagrams.
Such calculations are much simpler than the Monte Carlo simulations and, as we demonstrated, allow us to obtain temperature and field
variation of the ordering wave vectors, which are not accessible in the Monte Carlo  approach because the standard Metropolis algorithm does not
allow for measurement of  the free energy.

{\it Acknowledgements.}\ \  We thank Hiroyuki Nojiri for stimulating discussions and encouragement and appreciate helpful conversations with
Bj\"orn F\aa k and  Efim Kats. M.E.Z. acknowledges support by the exchange program of ICC-IMR, Tohoku University.

\end{document}